\documentstyle{mn}
\input epsf

\def\apj{ApJ}

\def\aj{AJ}
\def\aap{A\&\hskip-1pt A}

\def\mnras{MNRAS}

\def\lesssim{\mathrel{\hbox{\rlap{\hbox{\lower4pt\hbox{$\sim$}}}\hbox{$<$}}}}
\def\gtrsim{\mathrel{\hbox{\rlap{\hbox{\lower4pt\hbox{$\sim$}}}\hbox{$>$}}}}
\newcommand{\rvec}{\mbox{\boldmath $r$}}

\title[Spot-induced Anomalies in Caustic-crossing Binary Microlensing 
       Light Curves]
      {Variation of Spot-induced Anomalies in Caustic-crossing Binary
       Microlensing Event Light Curves}
\author[Chang \& Han]
{Heon-Young Chang$^1$ and Cheongho Han$^2$\thanks{e-mails:
hyc@ns.kias.re.kr (HYC); cheongho@astroph.chungbuk.ac.kr (CH)}\\
${}^1$Korea Institute for Advanced Study,
207-43 Cheongryangri-dong Dongdaemun-gu, Seoul 130-012, Korea\\
${}^2$Department of Physics, Chungbuk National University, Chongju
361-763, Korea}

\begin{document}
\maketitle
\vspace{-\abovedisplayskip}
\begin{abstract}
We investigate the pattern of anomalies in the light curves of 
caustic-crossing binary microlensing events induced by spot(s) on the lensed 
source star.  For this purpose, we perform simulations of events with various 
models of spots.  From these simulations, we find that the spot-induced 
anomalies take various forms depending on the physical state of spots, 
which is characterized by the surface brightness contrast, the size, the 
number, the umbra/penumbra structure, the shape, and the orientation with 
respect to the sweeping caustic.  We also examine the feasibility of 
distinguishing the two possibly degenerate types of anomalies caused by a 
spot and a transiting planet and find that the degeneracy in many cases can 
be broken from the characteristic multiple deviation feature in the 
spot-induced anomaly pattern caused by the multiplicity of spots.
\end{abstract}

\section{Introduction}

Microlensing experiments (MACHO: Alcock et al.\ 1993; EROS: Aubourg 1993)  
were originally proposed to search for Galactic dark matter in the form of 
massive compact halo objects by monitoring lensing-induced light variations 
of stars located in the Large Magellanic Cloud (Paczy\'nski 1986).  Besides 
this original goal, it was demonstrated that microlensing can also be applied 
to other fields of astronomy, especially stellar astrophysics (see the recent 
review of Gould 2001).  Over the last decade, this aspect of microlensing 
achieved important progress thanks partially to the theoretical studies on 
various methods to extract additional information about lensed source stars 
and more importantly to the large number of event detections from additional 
experiments directed towards the Galactic bulge (OGLE: Udalski et al.\ 1993; 
DUO: Alard \& Guibert 1997) and detailed light curves of events obtained 
from intensive followup observations (PLANET: Albrow et al.\ 1998; GMAN: 
Alcock et al.\ 1997; MPS: Rhie et al.\ 1999; MOA: Bond et al.\ 2001).

One of the lensing applications to stellar astrophysics is the detection 
and characterization of stellar spots.  Spot detection via microlensing 
is possible for high magnification events produced by the source's crossing 
of the lens caustic, which refers to the source position on which the 
magnification of a point source event becomes infinity.  For these events, 
one can resolve the source star surface because different parts of the source 
is magnified by different amounts due to the large gradient of magnification 
over the source during the caustic crossing (Gould 1994; Nemiroff \& 
Wickramasinghe 1994; Witt \& Mao 1994).  For a single lens, the location 
of the caustic is that of the lens itself (point caustic).  For a binary lens, 
the set of caustics forms closed curves in which each curve is composed of 
three or more concave line segments (fold caustic) that meet at cusps.  
Heyrovsk\'y \& Sasselov (2000) first investigated the possibility of lensing 
spot detections and showed that, for point-caustic-crossing single lens 
events, spots can cause fractional deviations in magnification larger than 
2 \%, which are detectable from followup observations.  Han et al.\ (2000) 
further investigated the possibility of detecting stellar spots from the 
observations of fold-caustic-crossing binary lens events and showed that the 
fractional deviations are comparable to those of point-caustic-crossing 
events.  However, these works were concentrated only on the possibility of 
spot detections, and thus no detailed investigation about the various forms 
of spot-induced anomalies in lensing light curves has been done.  Bryce \& 
Hendry (2000) mentioned some of the variations in their unpublished paper, 
but these analysis was only for point-caustic-crossing events, which are much 
less common than fold-caustic-crossing events.  In addition, the previous 
studies are based on very simplified assumptions  of spots and their host 
stars, e.g.\ a circular spot with a uniform surface brightness on also a 
uniform background stellar surface.  Therefore, extensive study about the 
microlensing signature of stellar spots for caustic-crossing binary lens 
events based on realistic models of spots and host stars is required.

Another reason for the necessity of studying spot-induced light curve 
deviations was recently raised by Lewis (2001).  He pointed out that if 
a source star possessing a planet is microlensed when the planet is 
transiting the source star surface, the resulting light curve will have 
similar deformation to those induced by a spot.  Then, unless one can 
distinguish the two types of deviations, analyses based on one naively 
adopted assumption for the cause of the deviation might result in false 
information about the source star.

In this paper, by performing realistic simulations of microlensing light 
curves of events occurred on spotted source stars, we investigate how the 
pattern of the spot-induced anomalies in lensing light curves varies for 
different states of spots.  Based on the results of this investigation, we 
also examine the feasibility of distinguishing the two types of anomalies 
caused by a spot and a transiting planet.

The paper is organized as follows.  In \S\ 2, we describe the basics of 
microlensing, which are  required to describe the lensing behavior of 
caustic-crossing binary lens events occurred on spotted source stars.  
In \S\ 3, by simulating lensing events occurred on source stars with various 
states of spots, we investigate the dependencies of the spot-induced anomaly 
pattern on various spot parameters, which characterize the physical state of 
spots.  In \S\ 4, we discuss the features of the spot-induced anomalies that 
can be used to distinguish the spot-induced anomalies from those caused by 
a transiting planet.  We conclude in \S\ 5.

\section{Basics of Microlensing}
If a point source located at $\rvec_{\rm S}$ on the projected plane of the 
sky is lensed by a binary lens system, where the individual lens components 
have masses of $m_i$ and projected locations of $\rvec_{{\rm L},i}$ ($i=1$ 
and 2), the positions of the resulting images, $\rvec_{\rm I}$, are obtained 
by solving the lens equation, which is expressed by
\begin{equation}
\rvec_{\rm S}\ =\ 
\rvec_{\rm I}\ -\ \theta_{\rm E}^{2}\  
\sum_{i=1}^2 {m_i\over M}\ 
{\rvec_{\rm I}-\rvec_{{\rm L},i} \over 
\left\vert\rvec_{\rm I}-\rvec_{{\rm L},i}\right\vert^2}\ ,
\end{equation}
where $M=m_1 + m_2$ are the total mass of the lens system and $\theta_{\rm E}$ 
represents the combined angular Einstein ring radius.  The Einstein ring radius 
is related to the total mass and the geometry of the lens system by
\begin{equation}
\theta_{\rm E}=
\left[  
{4GM\over c^2}
\left( {1\over D_{\rm OL}}-{1\over D_{\rm OS}} \right)
\right]^{1/2}\ ,
\end{equation}
where $D_{\rm OL}$ and $D_{\rm OS}$ are the distances to the lens and the 
source, respectively.  The magnifications of the individual images are given 
by the inverse of the Jacobian of the lens equation evaluated at each image
position $\rvec_{{\rm I},j}$;
\begin{equation}
A_j\ =\ 
\left({1\over \left\vert \det J \right\vert}\right)_{\rvec_{{\rm I},j}}\ ; \ \ \
\det J\ =\ \left\vert {\partial\rvec_{\rm S}\over\partial\rvec_{\rm I}}\right\vert\ ,
\end{equation}
and the total magnification is obtained by the sum of the magnifications of 
the individual images, i.e.\ $A=\sum_j^{N_{\rm I}} A_j$, where $N_{\rm I}$ is 
the total number of images.  The number of images is three when the source is 
outside the caustic curve.  As the source crosses the caustic, two additional 
images appear and thus the number of images becomes five when the source is 
inside the caustic curve.  Due to the infinite magnification during the 
source's crossing of the lens caustic, the light curves of a caustic-crossing 
binary lens event is characterized by sharp spikes.

In reality, the lensed source star is not a perfect point source, and thus 
the lensing magnification does not become infinity even for a caustic-crossing 
event.  For an event involved with an extended source, the magnification is 
given by the intensity-weighted magnification averaged over the source star 
surface, i.e.\
\begin{equation}
A_0 = {
\int\hskip-5pt\int I(\rvec) A(\rvec; \rvec_{{\rm L},i})\  d\Sigma_\ast
\over 
\int\hskip-5pt\int I(\rvec)\ d\Sigma_\ast}\ ,
\end{equation}
where $\rvec$ is the displacement vector of a point on the source surface 
with respect to the source center, $I(\rvec)$ is the surface brightness 
over the source star surface\footnote{Due to limb darkening, the surface 
brightness profile of a source star is not uniform but decreases with the 
increasing distance from the source center.}, $A(\rvec; \rvec_{{\rm L},i})$ 
is the point-source magnification at the point $\rvec$ caused by a binary 
lens with masses located at $\rvec_{{\rm L},i}\ (i=1\ {\rm and}\ 2)$, and 
the notations $\int\hskip-5pt\int \cdot\cdot\cdot\ d\Sigma_\ast$ represents 
the surface integral over the source star surface.  The existence of a spot 
(spots) causes further deviations in the lensing light curve.  For an event 
involved with a spotted source star, the lensing light curve is represented 
by
\begin{equation}
A_\bullet = {
\int\hskip-5pt\int \hskip-2pt I(\rvec) A(\rvec; \rvec_{{\rm L},i}) 
d\Sigma_\ast - 
\int\hskip-5pt\int \hskip-2pt f(\rvec') I(\rvec') A(\rvec'; \rvec_{{\rm L},i}) 
d\Sigma_\bullet
\over 
\int\hskip-5pt\int I(\rvec)\left[ 1-f(\rvec)\right]\ d\Sigma_\ast}\ ,
\end{equation}
where $\rvec'$ is the displacement vectors of a point on the spot with respect 
to the source center, $f(\rvec')$ represents the fractional decrement in the 
surface intensity due to the spot, and $\int\hskip-5pt\int \cdot\cdot\cdot\ 
d\Sigma_\bullet$ represents the surface integral over the spot region of the 
source star.

\begin{figure}
\epsfysize=8.5cm
\centerline{\epsfbox{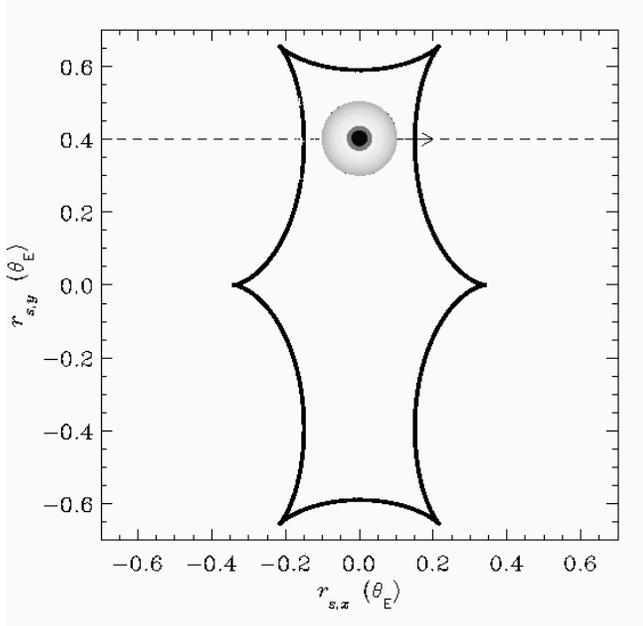}}
\vskip-0.4cm
\caption{
The source trajectory (dashed straight line with an arrow) and the caustics
(solid curves) of the binary lens system which are responsible for the tested 
events in the simulations.  The icon on the source trajectory visualizes the 
spotted source star.  The central small dark region symbolizes the spot 
(umbra region for the darkest area and penumbra region for the less dark 
area).  The circular region surrounding the spot represents the background 
stellar surface, where the outward increase of the grey scale symbolizes 
limb darkening.  The source star has an angular radius of $r_\ast=0.1
\theta_{\rm E}$.  The lens system has equal mass components separated by 
$\theta_{\rm E}$.  The coordinates are centered at the binary center and 
the abscissa coincides with the binary axis.
}
\end{figure}

\begin{figure} 
\epsfysize=9.0cm
\centerline{\epsfbox{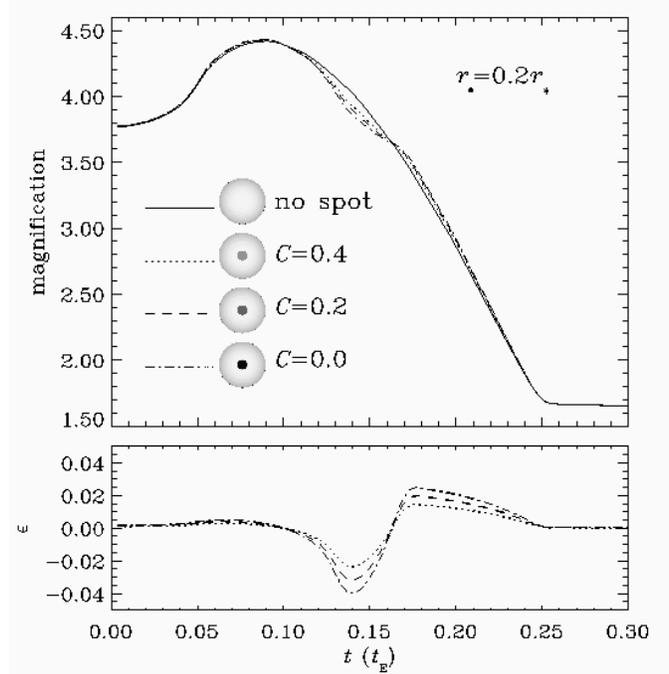}}
\caption{
Light curves of caustic-crossing binary lens events occurred
on source stars having spots with various contrast parameters.
The lower panel shows the changes of the fractional deviation from
the light curves of the unspotted source event (deviation curves).
}
\end{figure}

\section{Simulations}
In this section, we investigate the pattern of spot-induced anomalies in 
lensing light curves resulting from various physical states of spots.  For 
this, we perform simulations of events by using the formalism described in 
the previous section.  The details of the simulations are described below.

\subsection{Common Factors}
Since we are interested mainly in the dependency of the anomaly pattern on 
the spot parameters characterizing the physical states of spots, the tested 
events in the simulations are produced by a common binary lens system with 
a common source trajectory involved with also a common source star (unspotted
background part).  The lens system tested in the simulations has equal mass 
components separated by $\theta_{\rm E}$.  The source trajectory of the 
tested events is presented in Figure 1 along with the locations of the lens 
caustics.  The source star has an angular radius $r_\ast=0.1\theta_{\rm E}$ 
and its surface brightness profile is modeled by a linear form of
\begin{equation}
I_\nu (r) = 1-{\cal D}_\nu \left( 1-\sqrt{1-(r/r_\ast)^2} \right),
\end{equation}
where ${\cal D}_\nu$ is the limb-darkening coefficient whose value depends
on the observed band.  Assuming that events are observed both in $B$ and $I$ 
bands, the adopted values of the coefficients are ${\cal D}_B=0.912$ and 
${\cal D}_I=0.503$\footnote{We consider multi-band observations because 
the color information might be useful to break the possible degeneracy 
between the anomalies caused by a spot and a transiting planet, as mentioned 
by Lewis (2001), but we note that the light curves to be presented in the 
following subsection are those measured in $I$ band.}, which corresponds to 
those of a K giant star with $T_{\rm eff}=4750\ {\rm K}$, ${\rm log}\ g\sim 
2.0$, and a metallicity similar to the sun, i.e.\ $[{\rm Fe}/{\rm H}]\sim 0$ 
(Van Hamme 1993).  Since the spin rotation period of a giant star is much 
larger than the caustic sweeping time, we do not consider the effect of 
changing spot position during caustic crossings in our simulations.

\begin{figure}
\epsfysize=9.0cm
\centerline{\epsfbox{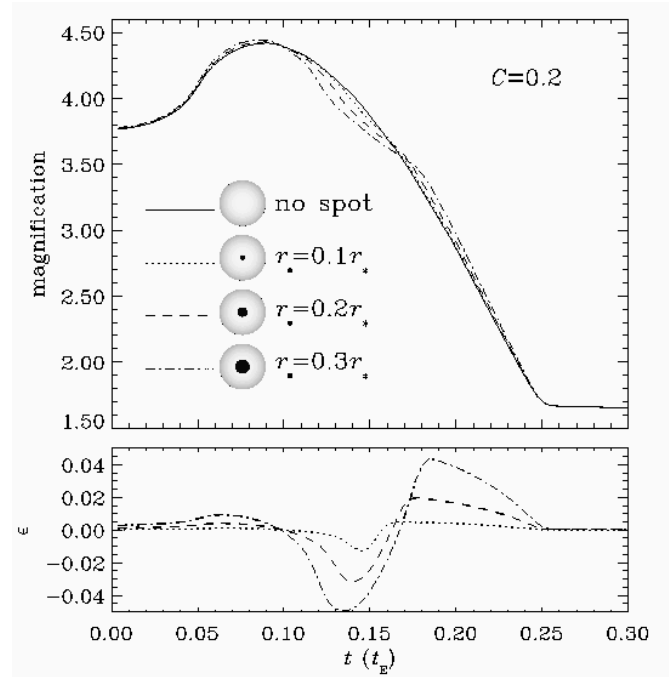}}
\caption{
Light and deviation curves of caustic-crossing binary lens events occurred 
on source stars having spots with various spot sizes.  
}
\end{figure}

\subsection{Spot Models and Resulting Anomaly Patterns}
The physical state of a spot is characterized by various factors.  These
factors include the size, the shape, and the surface brightness contrast 
with respect to the unspotted stellar region.  In addition, spots are likely 
to appear in groups and may have umbra/penumbra structures. To investigate 
the individual effects of these factors on the pattern of spot-induced 
anomalies in lensing light curves, in each set of simulations, we produce 
a series of light curves by varying one spot parameter but fixing other 
parameters.

First, we investigate the dependency of the spot-induced anomaly pattern 
on the surface brightness contrasts of spots.  In Figure 2, we present the 
light curves produced with spot models having different contrast parameters.  
The contrast parameter is defined by $C=I_\bullet/I_\ast$, where $I_\bullet$ 
and $I_\ast$ are the surface brightnesses of the source star at the spot 
position with and without the presence of the spot and it is related to the 
decrement factor $f$ in eq.\ (5) by $C=1/(1-f)$.  With this definition of 
the contrast parameter, the anomaly pattern induced by a transiting planet 
is produced by setting $C=0.0$, because the planet completely blocks the 
light from the source star.  In the simulations, we assume that the spots 
are located at the center of the source star and have a common radius of 
$r_\bullet=0.2r_\ast$ (and thus covers 4\% of the source star surface area).  
The time is expressed in units of the time scale that is required for the 
source to transit $\theta_{\rm E}$ (Einstein time scale, $t_{\rm E}$).  
To better show the deviation from the unspotted event light curve, we define 
a fractional magnification deviation by
\begin{equation}
\epsilon = {A_\bullet - A_0\over A_0},
\end{equation}
and its change as a function of time (deviation curve) is presented in the 
lower panel.  We test four different surface brightness contrast parameters 
of $C=1.0$ (no spot), 0.4, 0.2, and 0.0 (a completely dark spot or a transiting
planet).\footnote{For comparison, we note that the contrast parameter of a 
typical solar spot (umbra part) is $C\sim 0.21$ in $I$ band (Allen 1973).}  
We note that the deviation curve is asymmetric with respect to the time of 
source center's crossing of the caustic because not only the magnifications 
but also the gradients of the magnification on the left and right sides of 
the caustic are different each other.  We also note that the value of 
$\epsilon$ is larger than zero for some parts of the light curve due to the 
decrease in the source baseline flux caused by the spot, and thus smaller 
value of the denominator in eq.\ (5).  From the comparison of the light and 
deviation curves, one finds that darker spots, as expected, induce deeper 
hollows in the deviation curves.

\begin{figure}
\epsfysize=9.0cm
\centerline{\epsfbox{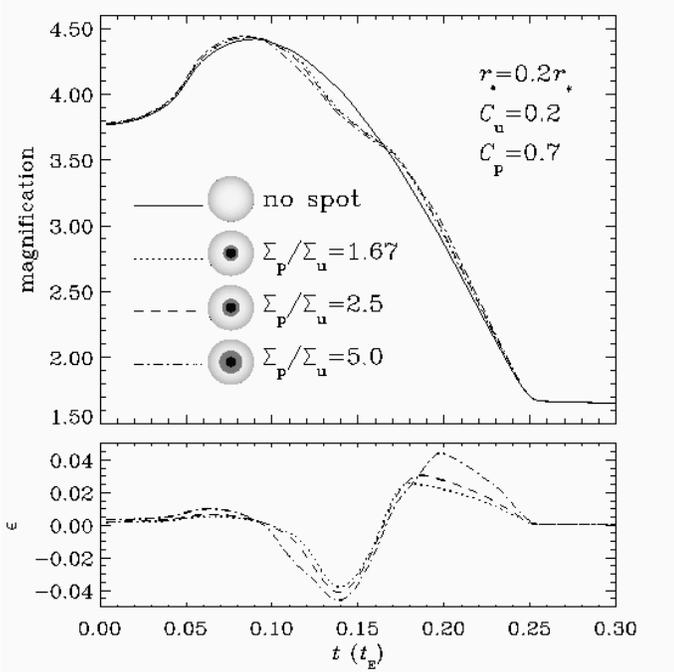}}
\caption{
Light and deviation curves of caustic-crossing binary lens events occurred 
on source stars having spots with various umbra/penumbra structures.  
}
\end{figure}

\begin{figure}
\epsfysize=9.0cm
\centerline{\epsfbox{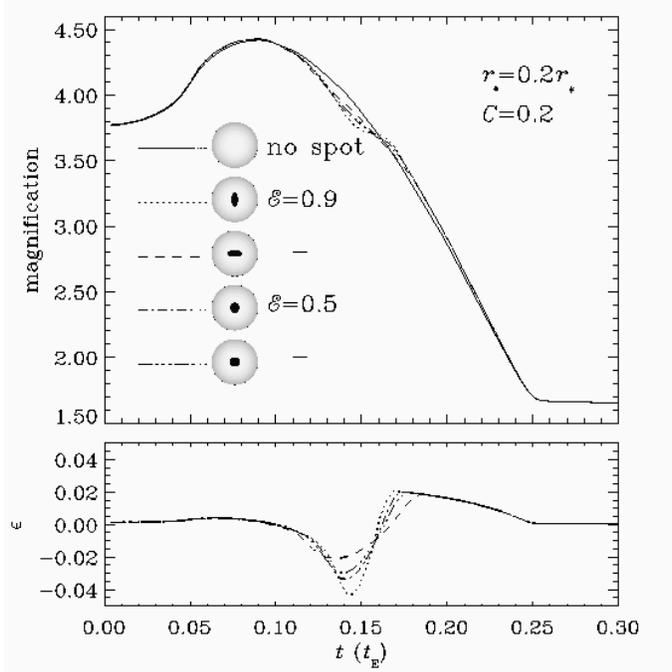}}
\caption{
Light and deviation curves of caustic-crossing binary lens events occurred 
on source stars having spots with various shapes and orientations with 
respect to the sweeping caustic.
}
\end{figure}

\begin{figure}
\epsfysize=9.0cm
\centerline{\epsfbox{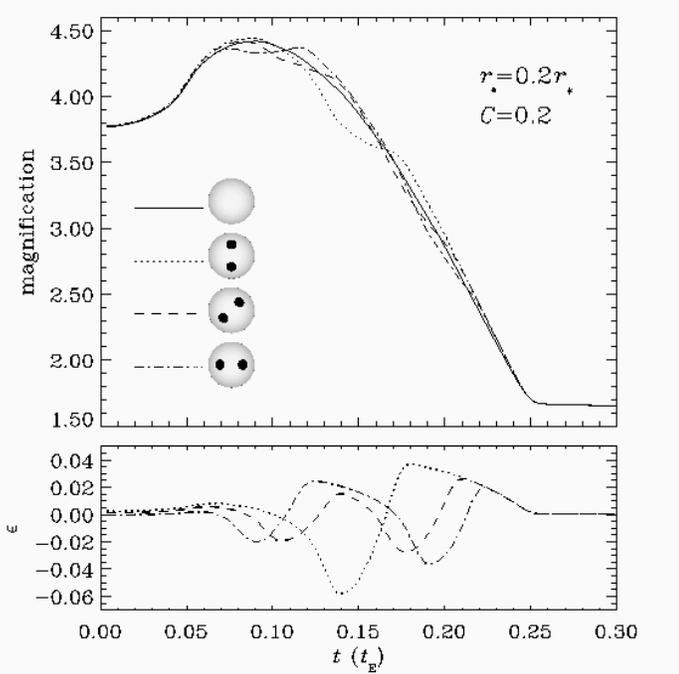}}
\caption{
Light and deviation curves of caustic-crossing binary lens events occurred 
on source stars having two spots.  The presented curves are for three 
different sets of the spot locations.
}
\end{figure}

Second, we investigate the dependency of the anomalies on the spot size.
Figure 3 shows the light curves produced with three different spot sizes 
of $r_\bullet=0.1r_\ast$, $0.2r_\ast$, and $0.3r_\ast$.  For all these 
events, we assume that the spots are located at the center of the source and 
have a common contrast parameter of $C=0.2$.  From the figure, one finds that 
as the spot size becomes bigger, the deviation curve becomes not only deeper 
but also broader.

Third, we probe the dependency of the anomaly pattern on the umbra/penumbra 
structure of spots.  We test three different spot structure models where the 
area of the umbra part, $\Sigma_{\rm u}$, is fixed while the area of the 
penumbra region has three different values of $\Sigma_{\rm p}=1.67 
\Sigma_{\rm u}$, $2.5 \Sigma_{\rm u}$, and $5.0 \Sigma_{\rm u}$.  In each 
model, we assume that the umbra region covers $4\%$ of the total source 
surface area.  The boundaries of the umbra and penumbra regions have a 
concentric circular shape centered at the center of the source star.  The 
adopted values of the contrast parameters are $C_{\rm u}=0.2$ and $C_{\rm p}
=0.7$ for the umbra and the penumbra parts, respectively.  One finds that 
the existence of the penumbra causes the deviation curve to become broader, 
but the increase in depth is less important due to the small surface 
brightness contrast of the penumbra region.  Although the curvature 
of the deviation curve changes when the caustic crosses the umbra/penumbra 
boundary (at around $t=0.12$ and 0.17 in the curve), the change is too small 
to be noticed.

Fourth, we investigate the anomaly pattern dependency on the shape of spots.  
The shape of a spot can differ from a circle either due to the projection 
effect or because of its intrinsic shape.  Moreover, spots tend to exist in 
a group and if small spots are closely located, the group as a whole will 
appear to be elongated.  We test two elliptical spots with eccentricities 
of $\varepsilon = \sqrt{a^2-b^2}/a=0.9$ and 0.5, where $a$ and $b$ are the 
semi-major and semi-minor axes of the ellipse, respectively.  We also 
investigate the effect of the spot orientation with respect to the sweeping 
caustic.  For this, we test two cases for each spot shape where the major 
axis of the spot is parallel with and normal to the fold caustic, respectively 
(see the spots on the icons of Figure 5).  The area of each elongated spot 
is normalized so that it is same as that of a circular spot with $r_\bullet
=0.2r_\ast$.  We assume that the surface brightness contrast is $C=0.2$.  
The resulting light curves are presented in Figure 5.  From the figure, one 
finds that the deviation curve becomes deeper and narrower when the elongated 
direction of the spot is parallel with the sweeping caustic.  This is because 
as the elongated direction of the spot aligns more with the caustic, larger 
part of the spot crosses the caustic, but during shorter period of time.

Finally, we investigate how the multiplicity of spots affects the anomaly 
pattern.  We test dual spot model, where the two spots have the same size, 
surface brightness contrast, and shape.  In Figure 6, we present the light 
curves resulting from three different sets of spot locations, which are 
marked on the corresponding icons.  We assume that the spots are intrinsically 
circular, but they are off-centered and thus appear to be elliptical.  The 
size of each spot is normalized such that the area is same as that of a 
circular spot with $r_\bullet=0.2r_\ast$.  We assume a surface contrast of 
$C=0.2$.  One finds that except for the case where the two spots are aligned 
with the caustic, the individual spots cause separate hollows in the 
deviation curve.

\section{Transiting Planets Versus Spots}
As mentioned, the lensing light curve anomalies induced by a transiting 
planet can mimic those induced by a spot.  However, there are several 
characteristics of a spot that make the spot differentiated from a transiting 
planet.  Then, for some of the spot-induced lensing light curve anomalies, 
it may be possible to isolate them from the anomalies induced by transiting 
planets.  One may think of several such distinctive characteristics of spots.  
First, unlike transiting planets, spots have in general non-zero surface 
brightness contrasts.  Second, spots normally have non-circular shapes, 
while the silhouettes of transiting planets are circular.  Third, spots are 
likely to have complex umbra/penumbra structures unlike the uniform contrast 
of transiting planets.  Fourth, spots would appear, by nature, in groups of 
two or more, while the chance for two planets simultaneously transit their 
host star is extremely low.  Based on the patterns of spot-induced anomalies 
presented in the previous section, however, one finds that the spot 
characteristics of the non-zero surface brightness, the non-circular shape, 
and the umbra/penumbra structure do not produce distinctive features in the 
anomaly pattern.  The major effects of these spot characteristics are making 
the width and the depth of the resulting deviation curve change, but these 
variations can be reproduced by modulating the planet size.  
On the contrary, the multiple deviation feature in the 
anomaly pattern caused by the multiplicity of spots is unique 
only for the spot-induced anomalies.  Since the
multiplicity is a general tendency for spots, we think that 
the spot-induced anomalies, in many cases, can be identified from 
this feature.

\begin{figure}
\vskip0.2cm
\epsfysize=8.1cm
\centerline{\epsfbox{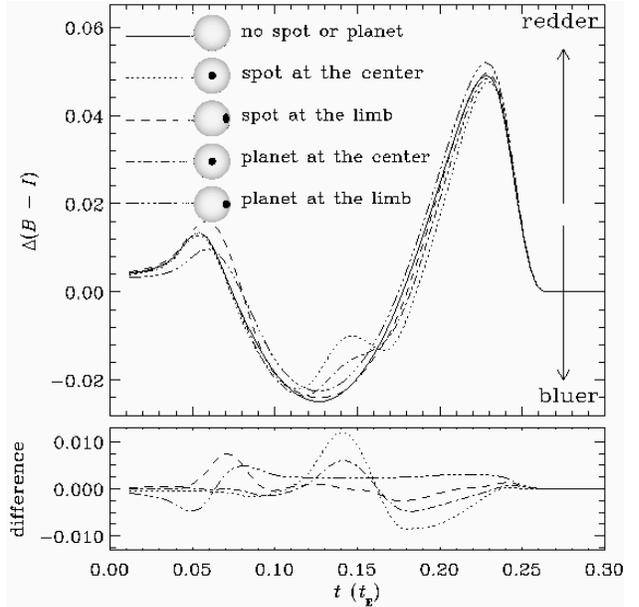}}
\caption{
Comparison of the color changes of caustic-crossing binary lens events 
occurred on source stars having a spot and a transiting planet.  For each 
case, we test two models where the spot (planet) is located at the center 
and the limb of the source star.  The lower panel shows the differences 
(in units of magnitude) of the color curves from that of the event produced 
without the spot or the planet (solid curve in the upper panel).
}
\end{figure}

Besides the anomaly pattern in lensing light curves, we additionally check 
the possibility to distinguish the two types of anomalies by using the color 
information obtained from multi-band photometry.  For this check, we simulate 
color changes of events occurred on source stars with a spot and a transiting 
planet.  For each case, we test two models where the spot (planet) is located 
at the center and the limb of the source star during the caustic crossing 
(see the icons in Figure 7).  We assume that both the spot and the planet 
covers 4\% of the source star surface area.  We also assume that observations 
are carried out in $B$ and $I$ bands.  We note that while the contrast 
parameter for the transiting planet is zero regardless of the observed band, 
the spot contrast parameter is  wavelength dependent.  The spot contrast 
parameter have different values depending on the observed wavelength band 
not only because the spot region of the stellar surface has a different 
temperature from that of the background stellar  surface but also because in 
different wavelengths one sees different layers of the stellar atmosphere 
(e.g.\ in longer wavelength, one sees cooler upper layer of the atmosphere).  
We assume that the spot contrast parameters are $C=0.05$ and 0.21 in $B$ and 
$I$ bands, respectively, by adopting those of a typical solar spot (Allen 
1973).  In Figure 7, we present the color change curves obtained from the 
simulations.  In th lower panel, we also present the difference (in units 
of magnitude) from the color change curve of the event produced without the 
spot or the planet.  We note that due to the limb darkening of the source 
star, color change occurs even without the presence of the spot or the planet 
on the source star.\footnote{For the description of the general color change  
patterns of both the point- and fold-caustic-crossing microlensing events 
occurred on limb-darkened source star, see Han \& Park (2001).} Therefore, 
the difference presented in the lower panel represents the additional color 
changes induced by the spot and the planet.  From the figure, one finds 
that the pattern of the additionally induced color changes takes various 
forms.  However, the amount of these additional color changes are expected 
to be very small ($\lesssim 0.01$ mag), and thus we think that color 
information will be useful only when the the uncertainty in the measured 
color is very small.

\section{Conclusion}
We have investigated the patterns of spot-induced anomalies in 
caustic-crossing binary lens events.  For this purpose, we performed 
simulations of events with various models of the physical state of spots 
which is characterized by the surface brightness contrast, the size, the 
number, the umbra/penumbra structure, the shape, and the orientation with 
respect to the sweeping caustic.  From these simulations, we learned that 
the spot-induced anomalies take various forms depending on these factors.  
We also examined the feasibility of distinguishing the two possibly 
degenerate types of anomalies caused by a spot and a transiting planet
and found that the degeneracy in many cases can be broken from the
characteristic multiple deviation feature in the spot-induced anomaly 
pattern caused by the multiplicity of spots.

\bigskip
This work was supported by a grant (2001-DS0074) from Korea Research 
Foundation (KRF).

{}

\end{document}